\pdfoutput=1
\documentclass[10pt,conference,final]{IEEEtran}

\newcommand{\e}{\mathrm{e}}

\usepackage{cite}
\usepackage{url}
\usepackage{amsmath}
\usepackage{amssymb}
\usepackage{amsthm}
\usepackage{graphicx}
\usepackage{xcolor}
\usepackage{dsfont}

\makeatletter
\let\MYcaption\@makecaption
\makeatother

\usepackage{caption}
\usepackage[font=footnotesize]{subcaption}

\makeatletter
\let\@makecaption\MYcaption
\makeatother

\newtheorem{proposition}{Proposition}
\newtheorem{lemma}{Lemma}
\theoremstyle{definition}
\newtheorem*{replicatrick}{Replica trick}
\theoremstyle{remark}

\newcommand{\E}{\mathsf{E}}

\newcommand{\R}{\mathbb{R}}
\newcommand{\indF}{\mathds{1}}
\newcommand{\setX}{\mathcal{X}}
\newcommand{\vm}[1]{\boldsymbol{#1}}
\newcommand{\trans}{\mathsf{T}}

\newcommand{\dx}{\mathrm{d}}
\newcommand{\Dx}{\mathrm{D}}
\newcommand{\extr}{\mathop{\rm extr\/}}

\newcommand{\argmin}{\mathop{\rm arg\hspace{2pt}min\/}}

\newcommand{\mse}{\mathsf{mse}}
\newcommand{\ncostF}{C}

\newcommand{\M}{M}
\newcommand{\N}{N}

\newcommand{\Ap}{A}
\newcommand{\Bp}{\chihat}

\newcommand{\NR}{u}

\newcommand{\Qhat}{\hat{Q}}
\newcommand{\mhat}{\hat{m}}
\newcommand{\chihat}{\hat{\chi}}

\newcommand{\varnoise}{\sigma^{2}_{w}}

\begin{document}

\title{Statistical Mechanics Approach\\ to Sparse Noise Denoising}

\author{\IEEEauthorblockN{Mikko Vehkaper{\"a}$^{1,2}$, Yoshiyuki Kabashima$^{3}$, Saikat Chatterjee$^{1}$} 
\IEEEauthorblockA{$^{1}$%
KTH Royal Institute of Technology and the ACCESS Linnaeus Center, 
SE-100 44, Stockholm, Sweden}
\IEEEauthorblockA{$^{2}$%
Aalto University, P.O. Box 11000, FI-00076 AALTO, Finland
}%
\IEEEauthorblockA{$^{3}$%
Tokyo Institute of Technology, Yokohama, 226-8502, Japan
}%
E-mails: \url{mikkov@kth.se}, \url{kaba@dis.titech.ac.jp},
\url{sach@kth.se}}

\maketitle

\begin{abstract}
Reconstruction fidelity of sparse signals contaminated 
by sparse noise is considered.  Statistical mechanics
inspired tools are used to show that the 
$\ell_{1}$-norm based convex optimization algorithm exhibits 
a phase transition between the possibility 
of perfect and imperfect reconstruction.  
Conditions characterizing 
this threshold are derived and the mean
square error of the estimate is obtained
for the case when perfect reconstruction is not possible.  
Detailed calculations are provided to expose the 
mathematical tools to a wide audience.
\end{abstract}

\section{Introduction}
\label{sec:intro}

Sparse signal estimation for linear underdetermined systems 
has attracted wide interest in signal processing community during
the recent years. This is not surprising since the 
general class of sparse problems is encountered in many 
applications, such as, linear regression \cite{Miller_2002_Subset_Selection_in_Regression}, 
multimedia \cite{Daudet_Molecular_Matching_Pursuit,
Sastry_2009_Face_recognition}, 
and compressive sampling (CS) \cite{Donoho_2006_Compressed_sensing,CS_introduction_Candes_Wakin_2008},
to name just a few.

The present paper considers a CS setup where the sparse vector 
$\vm{x} \in \R^{\N}$ is observed via noisy linear measurements
\begin{equation}
\vm{y} = \vm{A} \vm{x} + \vm{w},
\label{eq:Sparse_Representation_with_Noise}
\end{equation}
where
$\vm{A} \in \R^{M \times N}$ represents the compressive
$(M<N)$ sampling system 
and $\vm{y} \in \R^{M}$ is the observed vector.
The measurement errors are captured by the 
additive noise vector 
$\vm{w} \in \R^{M}$.
The task is to reconstruct $\vm{x}$ from $\vm{y}$, given  $\vm{A}$,
but without detailed information about the statistics 
of $\vm{x}$ and $\vm{w}$.

A prominent approach for finding a sparse solution to
\eqref{eq:Sparse_Representation_with_Noise} is by solving
a (convex) optimization problem of the form
\begin{equation}
\hat{\vm{x}}_{\lambda} = 
\argmin_{\vm{x} \in \mathbb{R}^{N}} \hspace{3pt} 
\left\{\ncostF_{\vm{y},\vm{A}}(\vm{x})
+ \lambda \| \vm{x} \|_{1}
\right\},
\label{eq:CS_Standard_Problem_LASSO}
\end{equation}
where $\|\vm{x}\|_{1} = \sum_{n} |x_{n}|$. 
The cost function
$\ncostF_{\vm{y},\vm{A}}(\vm{x}) \geq 0$,
that may depend on the realizations of 
$\vm{y}$ and $\vm{A}$, is typically
chosen so that \eqref{eq:CS_Standard_Problem_LASSO}
can be obtained using convex optimization tools like
\texttt{cvx} \cite{cvx}.
In addition to the choice of $\ncostF_{\vm{y},\vm{A}}(\vm{x})$, 
the solution also depends on the 
\emph{regularization parameter} $\lambda$.  In general, 
finding the optimal value
of $\lambda$ is not a trivial task.

For the case of (dense) Gaussian noise, 
the standard approach is to set 
$\ncostF_{\vm{y},\vm{A}}(\vm{x}) = 
\| \vm{y} -  \vm{A} \vm{x} \|^{2}_{2}$,
reducing \eqref{eq:CS_Standard_Problem_LASSO} to 
the so-called LASSO estimator \cite{Tibshirani_1996_lasso}.
For non-zero noise variance,
the solution obtained through LASSO is not exact,
but if the noise has some structure, like sparsity,
perfect reconstruction may again be feasible \cite{5484996}.
Some \emph{applications where sparse noise} can
be encountered are:
impulsive noise \cite{Carrillo_impulsive_noise_2010}, 
salt-and-pepper noise in an image, a sensor scenario 
where few measurements are corrupted but the other ones 
are good \cite{Laska_Sparse_Noise_2009}, and 
dictionary learning with sparse noise \cite{Sparse_Noise_for_Dictionary_Learning_2011}. 

Let us consider a setup similar to \cite{5484996}, 
where both $\vm{x}$ and $\vm{w}$ are sparse, and 
the cost function is chosen as
\begin{equation}
\label{eq:cost_l1}
\ncostF_{\vm{y},\vm{A}}(\vm{x}) = 
\| \vm{y} -  \vm{A} \vm{x} \|_{1},
\end{equation}
to guarantee that 
\eqref{eq:CS_Standard_Problem_LASSO}
is a convex optimization problem.
Then, \emph{we ask the following questions}:
\begin{enumerate}
\item
Given that the signal and noise are sparse 
and convex optimization based on 
\eqref{eq:CS_Standard_Problem_LASSO} and \eqref{eq:cost_l1} is used 
for reconstruction,
what \emph{compression ratios}  $\alpha = \M / \N$ allow
a perfect reconstruction of $\vm{x}$?
\item
What is the mean square error (MSE)
of the sparse estimate of 
$\vm{x}$ outside of this region?
\end{enumerate}
We answer these questions
in the large system limit (LSL) and report the 
\emph{sharp threshold} for $\alpha$ that separates the two phases 
of reconstruction fidelity.
The key technique is the \emph{replica method}%
\footnote{Drawback of the replica method 
is that some of its steps are still lacking formal proof.
Hence, it can be considered to be at most a ``semi-rigorous'' 
analytical tool.  It is, however, routinely 
used in equilibrium statistical mechanics and its predictions 
are often verified by experiments. } 
developed in equilibrium statistical mechanics,
where it is used to study large-scale behavior of 
disordered physical systems, such as, spin glasses.
It has also been used in information theory
\cite{Tanaka-2002, Guo-Verdu-2005,
Zaidel-rsb-2012,
Takeuchi-etal-trit2012} and 
CS \cite{Kabashima-Wadayama-Tanaka-2009,
Kabashima-Vehkapera-Chatterjee-2012,
Rangan-Fletcher-Goyal-2012,
Tulino-etal-2013}, where quantities like mutual information and
MSE play the role of thermodynamic variables.

\section{Problem Formulation and Methods}
\label{sec:problem_formulation}

Consider the set of noisy measurements
\eqref{eq:Sparse_Representation_with_Noise} and assume
that \emph{both the signal and the noise are sparse} random vectors
(RVs). 
Let us define a parametrized mixture distribution 
\begin{equation}
p(z;\,\rho,\sigma^{2}) = 
(1-\rho) \delta(z) +
\rho g_{z}(0,\sigma^{2}),
\label{eq:mix_dist}
\end{equation}
where 
$g_{z}(\mu,\sigma^{2})=
\e^{-(z-\mu)^{2}/2\sigma^{2}}/\sqrt{2\pi \sigma^{2}}$,
and
$\delta(z)$ is the Dirac delta function.
Let the elements of $\vm{x}$ (resp.\ $\vm{w}$) be
independently and identically distributed (IID) according 
to $p(x;\,\rho_{x},\sigma_{x}^{2})$ (resp.\ 
$p(w;\,\rho_{w},\sigma_{w}^{2})$).  The
$\rho \in [0,1]$ in \eqref{eq:mix_dist} is 
the fraction of non-zero elements in the vector and
$\sigma^{2}$ is their variance. 
The measurement process is taken to be random so that
the elements of $\vm{A}$ are IID with 
density $g_{A}(0,1 / \N)$.
The ratio between the 
number of observables and the unknown parameters 
is denoted $\alpha = \M/\N$.  
To use statistical mechanics tools,
we next write the problem 
in a probabilistic framework.

Let us consider the optimization
problem \eqref{eq:CS_Standard_Problem_LASSO}
with the $\ell_{1}$-cost \eqref{eq:cost_l1}.
Assume the system is in the LSL  
$\M,\N\to\infty$, where the
compression ratio $\alpha = \M /\N$
and the density of the signal and noise
$\rho_{x},\rho_{w}$ remain as finite constants.
Let the postulated prior of $\vm{x}$ be
proportional to the Laplace distribution, namely,
$q_{\beta,\lambda} (\vm{x}) \propto
\e^{-\beta\lambda\|\vm{x}\|_{1}},$
where $\beta \geq 0$.
The postulated distribution of
the measurement process has
the same form, that is,
$q_{\beta} (\vm{y} \mid \vm{A}, \vm{x})
\propto \e^{-\beta
\| \vm{y} - \vm{A} \vm{x}\|_{1}},$
and the (mismatched) conditional 
mean estimator of $\vm{x}$ reads by
definition
\begin{align}
\label{eq:pme_invtemp_beta}
\left\langle \vm{x}; \lambda \right \rangle_{\beta}
&=
\frac{1}{Z_{\beta}(\vm{y},\vm{A};  \lambda)}
\int
\vm{x}
q_{\beta} (\vm{y} \mid \vm{A}, \vm{x})
q_{\beta,\lambda} (\vm{x}) \mathrm{d} \vm{x},
\end{align}
where 
$Z_{\beta}(\vm{y},\vm{A};  \lambda)
=\int \e^{-\beta (
\| \vm{y} - \vm{A} \vm{x}\|_{1}
+ \lambda \|\vm{x}\|_{1})} \mathrm{d} \vm{x}$. 
Then, the zero temperature estimate
$\hat{\vm{x}}_{\lambda} = 
\langle \vm{x}; \lambda \rangle_{\beta\to\infty},$
is the solution to the 
original optimization problem defined by 
\eqref{eq:CS_Standard_Problem_LASSO} and
\eqref{eq:cost_l1}.

\subsection{Replica Method}
\label{sec:statmech_form2}

The key for finding the statistical properties 
of the reconstruction \eqref{eq:pme_invtemp_beta}
is the normalization factor or \emph{partition function}
$Z_{\beta}(\vm{y},\vm{A};  \lambda)$.
Based on the statistical
mechanics approach, our goal is to assess the 
\emph{free energy}
$f_{\beta}(\vm{y},\vm{A}; \lambda) 
= -\frac{1}{\beta \N} \ln Z_{\beta}(\vm{y},\vm{A}; \lambda)$,
when $\N\to\infty$
and obtain the desired statistical properties from it.
This is, however, difficult since 
$f_{\beta}$ depends on the observations
and the measurement process.  
If the averaged quantity
$f_{\beta}(\lambda) =
\E  f_{\beta}(\vm{y},\vm{A}; \lambda)$ is considered instead, 
a new problem arises in assessing the expectation over  logarithm.
We may reformulate the problem 
by writing 
\begin{equation}
\label{eq:freeE_replica_real}
f(\lambda) = 
-\lim_{\beta,\N\to\infty} \frac{1}{\beta \N}
\lim_{\NR\to 0^{+}}
\frac{\partial}{\partial \NR}
\ln \E \{[Z_{\beta}(\vm{y},\vm{A};  \lambda)]^{\NR}\},
\end{equation}
and remark that so-far the development has been rigorous.
Unfortunately, obtaining an expression for
\eqref{eq:freeE_replica_real} is still difficult so
we resort to the replica trick in order to proceed.

\begin{replicatrick}
Consider the free energy in
\eqref{eq:freeE_replica_real}.  Assume that the
limits commute, which in conjunction with the expression
\begin{equation}
\label{eq:Z_replicated_1}
[Z_{\beta}(\vm{y},\vm{A};  \lambda)]^{\NR}
=
\int 
\prod_{a=1}^{\NR}
\e^{-\beta (
\| \vm{y} - \vm{A} \vm{x}^{a}\|_{1}
+ \lambda \|\vm{x}^{a}\|_{1})} \mathrm{d} \vm{x}^{a}
\end{equation}
for $u =1,2,\ldots$ allows the evaluation of the expectation in
\eqref{eq:freeE_replica_real}
as a function of $u \in \R$. The functional expression is utilized in taking the limit of $u \to 0^+$. 
\end{replicatrick}

The assumption that the variable $\NR$ 
(number of replicas) can be first treated as a 
non-negative integer and then extended to the set of real numbers
has no rigorous mathematical proof in general.
The predictions of the replica method, however,
tend to be accurate when compared to 
experiments.

The general scheme of the following analysis consists of
first assessing \eqref{eq:freeE_replica_real} using the 
replica trick and then identify the 
parameters that describe the MSE of the reconstruction.
Finally, requiring that the MSE vanishes provides the 
threshold for perfect recovery.
The next section reports
the outcomes of the analysis and Section~\ref{sec:replicas}
contains the derivations.

\section{Results and Discussion}

Let $Q$ denote the standard Q-function and define
\begin{IEEEeqnarray}{rCl}
s(x) &=& \frac{1}{x^{2}}
\big[1 - 2 Q(x)\big]
    - \sqrt{\frac{2}{\pi x^{2} }}
       \e^{-\frac{x^{2}}{2}},
\label{eq:s-func_prop} \\
r_{ \lambda}(h) &=& \lambda\sqrt{\frac{h}{2\pi}} 
\e^{-\frac{\lambda^{2}}{2 h}} -
(\lambda^{2}+h)Q\bigg(\frac{\lambda}{\sqrt{h}}\bigg).
\label{eq:rfunc_prop}
\end{IEEEeqnarray}
Then, under the (technical) assumption of 
\emph{replica symmetric ansatz}
(see Section~\ref{sec:replicas} for definition and \cite{Tanaka-2002,Guo-Verdu-2005,Zaidel-rsb-2012} for 
further discussion),
the following results are obtained.

\begin{proposition}
\label{prop:perf_reconstruction}
Fix $\lambda,\alpha,\rho_{x},\rho_{w}$ 
and let the variances 
 $\sigma_{x}^{2}$ and $\sigma_{w}^{2}$
be finite and non-zero.  
Then, the critical threshold for the perfect reconstruction,
$\mse \to 0$,
is given by the solution of
\begin{IEEEeqnarray}{rCl}
\label{eq:A_success_prop}
\Ap &=&
   \frac{%
   \rho_{x}(\lambda^{2}+\Bp) - 2 (1-\rho_{x}) r_{\lambda}(\Bp)}
   {%
    \big[ 2(1-\rho_{x})
   Q\big(\lambda/\sqrt{\Bp}\big)
   +  \rho_{x}\big]^{2}}, \\
\Bp &=&
\alpha(1-\rho_{w})
\bigg\{\Ap
\bigg[1 - 2 Q\bigg(\frac{1}{\sqrt{\Ap}}\bigg)\bigg]
    - \sqrt{\frac{2 \Ap}{\pi}}
      \e^{-1/(2 \Ap)}
  \IEEEeqnarraynumspace\nonumber\\
 && \qquad \qquad \qquad + 2Q\bigg(\frac{1}{\sqrt{\Ap}}\bigg)
\bigg\}
+\alpha\rho_{w}, 
\label{eq:chihat_success_prop} 
\end{IEEEeqnarray}
that satisfies the condition
\begin{equation}
\label{eq:success_condition_prop}
\alpha (1-\rho_{w})
    \Big[1 - 2 Q\Big(\frac{1}{\sqrt{\Ap}}\Big)\Big]
 =
(1-\rho_{x})
 2 Q\Big(\frac{\lambda}{\sqrt{\Bp}}\Big)
 +  \rho_{x}.
\end{equation}
The solution can be found by numerically iterating 
\eqref{eq:A_success_prop} and \eqref{eq:chihat_success_prop} 
until convergence and then checking if \eqref{eq:success_condition_prop} 
holds.
\end{proposition}

The above result
gives the \emph{critical threshold for the 
compression ratio} 
$\alpha_{\mathrm{c}}(\lambda,\rho_{x},\rho_{w})$
that guarantees vanishing 
MSE of reconstruction.  More precisely, if
$\alpha_{\mathrm{c}}(\lambda,\rho_{x},\rho_{w})$ is a solution to 
Proposition~\ref{prop:perf_reconstruction}, then for all
$\alpha < \alpha_{\mathrm{c}}(\lambda,\rho_{x},\rho_{w})$ 
we have perfect reconstruction
in the MSE sense, while 
$\alpha > \alpha_{\mathrm{c}}(\lambda,\rho_{x},\rho_{w})$ leads
to non-vanishing MSE.  
Note that the threshold depends on the
regularization parameter $\lambda$ 
and densities of the source and noise 
$\{\rho_{x},\rho_{w}\}$, but is independent of the 
variances of the non-zero elements of signal $\sigma_{x}^{2}$ and
noise $\sigma_{w}^{2}$.  
With Proposition~\ref{prop:perf_reconstruction}, we have 
thus answered the first question laid out 
in Section~\ref{sec:intro}.

\begin{proposition}
Let the system be outside of the 
perfect reconstruction phase given by 
Proposition~\ref{prop:perf_reconstruction}, i.e.,
the compression ratio is above the threshold
$\alpha > \alpha_{\mathrm{c}}(\lambda,\rho_{x},\rho_{w})$.
The MSE of the sparse signal estimate 
obtained with
\eqref{eq:CS_Standard_Problem_LASSO}~and~\eqref{eq:cost_l1} 
is then
\begin{IEEEeqnarray}{rCl}
\mse &=& \rho_{x}\sigma_{x}^{2}
- 4\sigma^{2}_{x}\rho_{x}
Q\bigg(\frac{\lambda}{\sqrt{\chihat+\sigma_{x}^{2}\mhat^{2}}}\bigg)
\nonumber\\
&& - 2 \mhat^{-2} \big[
  (1-\rho_{x}) r_{\lambda}(\chihat)
+ \rho_{x} 
r_{\lambda}(\chihat+\sigma_{x}^{2}\mhat^{2}) \big],
\IEEEeqnarraynumspace
\label{eq:mse_prop_1}
\end{IEEEeqnarray}
where the required parameters can be obtained by solving 
the following set of coupled equations
\begin{IEEEeqnarray}{rCl}
\chi &=&
 \frac{2}{{\mhat}} 
 \bigg[
  (1-\rho_{x}) 
Q\bigg(\frac{\lambda}{\sqrt{\chihat}}\bigg)
+ \rho_{x} 
Q\bigg
(\frac{\lambda}{\sqrt{\chihat+\sigma_{x}^{2}\mhat^{2}}}
\bigg)\bigg],
\label{eq:chi_prop} \IEEEeqnarraynumspace\\
\hat{m} &=&
\frac{\alpha(1-\rho_{w})}{\chi}
\bigg[1 - 2 Q\bigg(\frac{\chi}{\sqrt{\mse}}\bigg)\bigg]
\nonumber\\
&&+
\frac{\alpha\rho_{w}}{\chi}
\bigg[1 
- 2 Q\bigg(\frac{\chi}{\sqrt{\mse+\varnoise}}\bigg) \bigg],
\label{eq:hatQ-hatm_prop} \\
\chihat &=&
\alpha(1-\rho_{w}) \bigg[
s\bigg(\frac{\chi}{\sqrt{\mse}}\bigg) 
+2Q\bigg(\frac{\chi}{\sqrt{\mse}}\bigg)
\bigg] \nonumber\\
&&+\alpha\rho_{w}\bigg[ 
s\bigg(\frac{\chi}{\sqrt{\mse+\varnoise}}\bigg)
\!+\! 2 Q\bigg(\frac{\chi}{\sqrt{\mse+\varnoise}}\bigg)
\bigg].
\label{eq:chihat_prop} \IEEEeqnarraynumspace 
\end{IEEEeqnarray}
The solution can be found by numerically iterating the equations 
until convergence is reached.
\label{prop:mse}
\end{proposition}

With Proposition~\ref{prop:mse} we have answered 
the second question in Section~\ref{sec:intro}, namely, 
how does the MSE behave when perfect reconstruction is not possible.
It is important to note that Proposition~\ref{prop:mse} reduces
to Proposition~\ref{prop:perf_reconstruction} when we enforce
the condition $\mse\to 0$.  Taking the limit is, however, somewhat
subtle as explained in Section~\ref{sec:replicas}.
Note that in principle, one could observe the vanishing MSE also by setting 
$\alpha < \alpha_{\mathrm{c}}(\lambda,\rho_{x},\rho_{w})$
and numerically evaluating 
\eqref{eq:mse_prop_1}~--~\eqref{eq:chihat_prop}.  Some numerical 
difficulties, however, arise in this case 
since $\mhat\to\infty$ and $\chi\to 0$
holds for perfect reconstruction.
\begin{figure}[h]
        \centering
        \begin{subfigure}[t]{\columnwidth}
                \centering
                \includegraphics[width=\columnwidth]{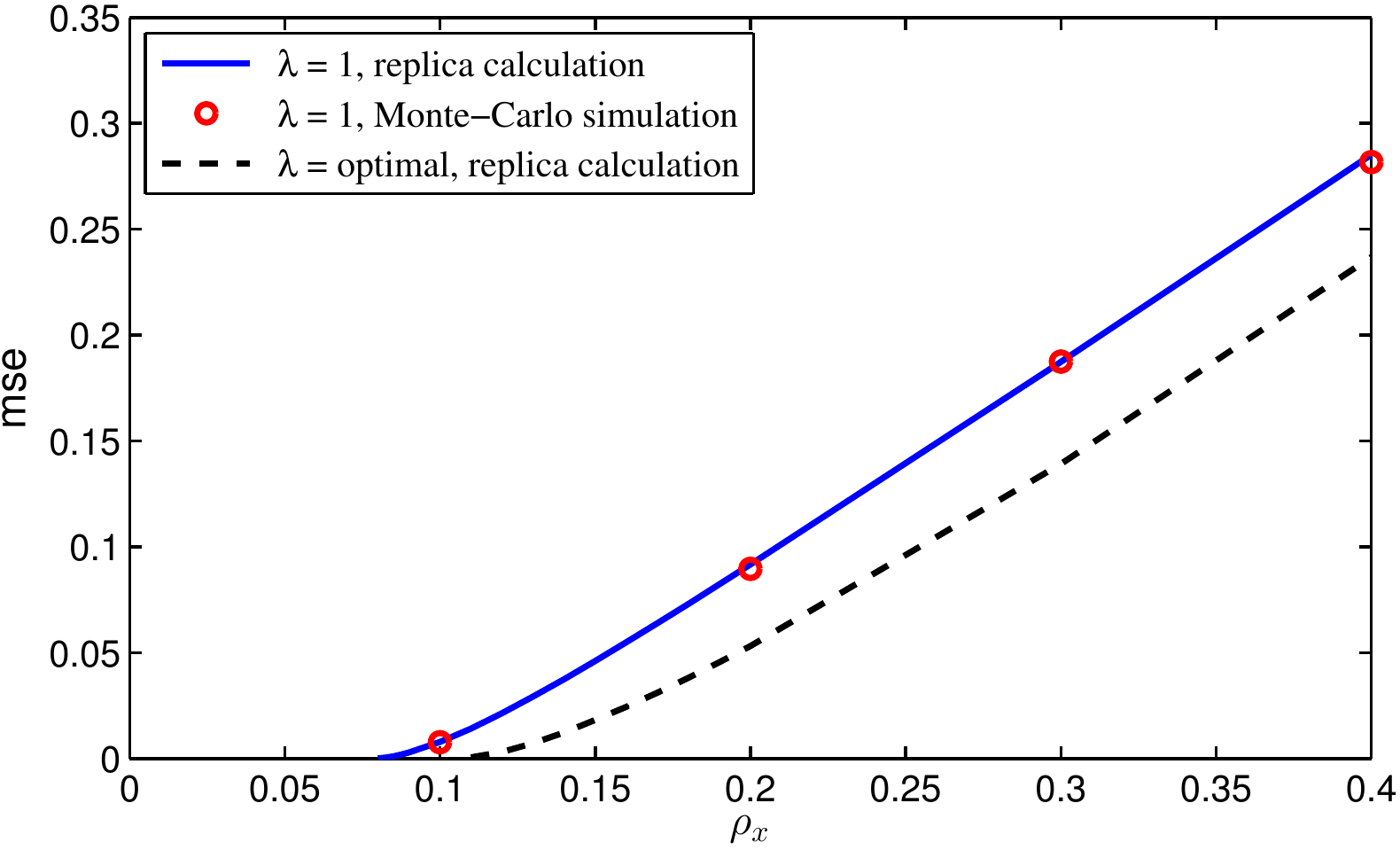}
                \vspace*{-5mm}
                \caption{
                MSE of reconstruction for
                $\alpha = 1/2$, $\rho_{w} = 0.1$ 
                and $\sigma^{2}_{x} = \sigma^{2}_{w}.$  
                Lines for 
                replica analysis based results and 
                markers for simulations.}
                \label{fig:Fig1_1}
        \end{subfigure}
        
        \begin{subfigure}[t]{\columnwidth}
        \vspace*{3mm}
                \centering
                \includegraphics[width=\columnwidth]{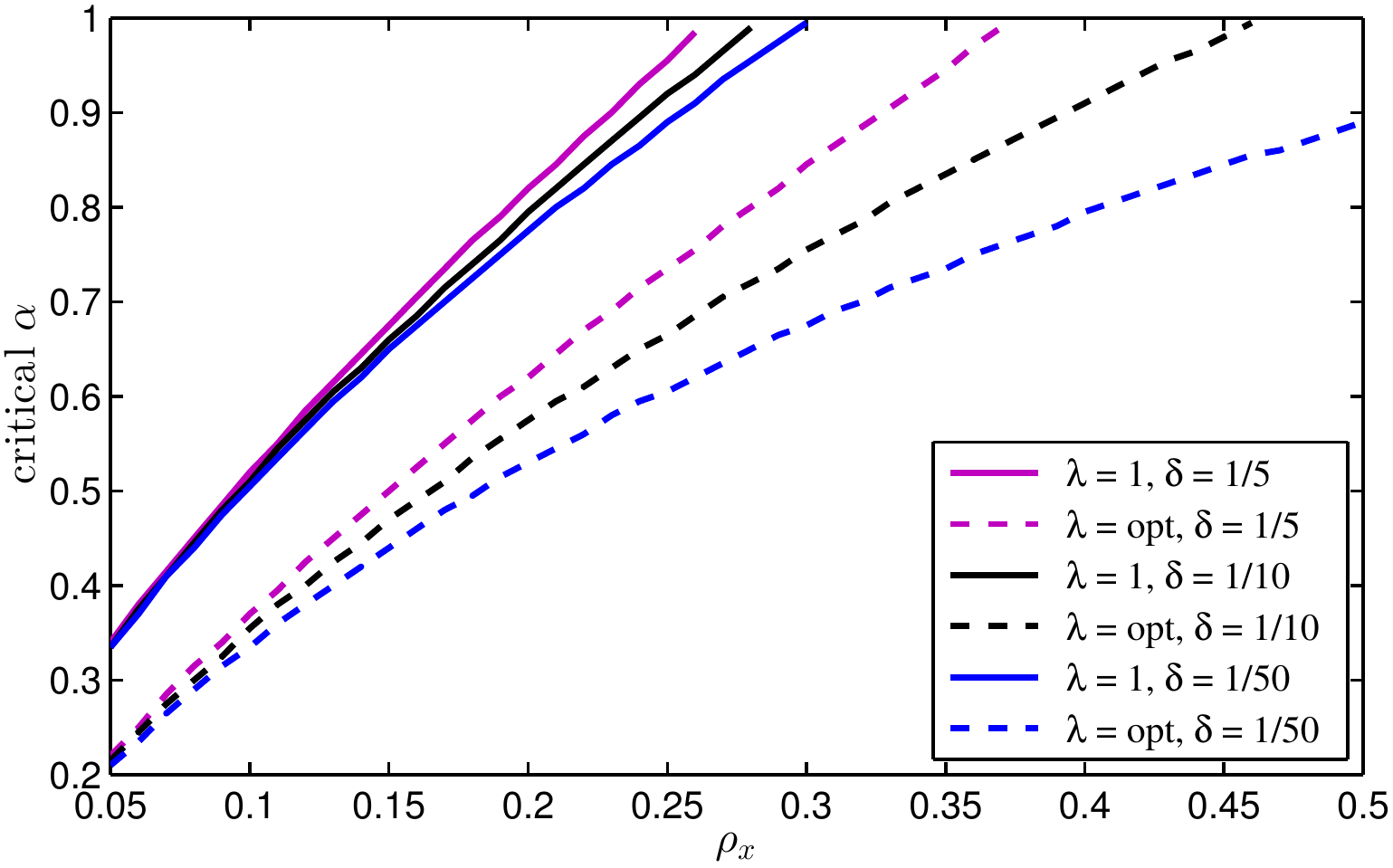}
                \vspace*{-5mm}
                \caption{Replica method based critical condition for perfect reconstruction.  Solid lines for $\lambda=1$
                and dashed lines for optimal $\lambda$.}
                \label{fig:Fig1_2}
        \end{subfigure}
        \vspace*{-1mm}
        \caption{Reconstruction performance vs.\ 
        signal density $\rho_{x}$.}
        \vspace*{1mm}
\end{figure}

Mean square error predicted by
Proposition~\ref{prop:mse}
is shown in Fig.~\ref{fig:Fig1_1}.  
Numerical experiments obtained with 
\texttt{cvx} \cite{cvx} are also given.
Below the thresholds 
$\rho_{x} = 0.0770$ and 
$\rho_{x} = 0.1030$ for $\lambda = 1$
and $\lambda = \text{optimal}$, respectively, 
the MSE of the reconstruction vanishes.
Figure~\ref{fig:Fig1_2} shows the effect of 
$\lambda$ on the perfect recovery threshold 
given in Proposition~\ref{prop:perf_reconstruction}.
Here $\rho_{w} = \delta \rho_{x}$, where
$\delta = 1/5, 1/10, 1/50$.  For given $\rho_{x}$
we find the critical threshold $\alpha_{\mathrm{c}}(\lambda,\rho_{x},\rho_{w})$ that admits 
perfect reconstruction, so that
the MSE vanishes for the set of parameters that 
lie above the selected curve.  The results 
demonstrate that the choice of the regularization
parameter $\lambda$ has 
a significant impact on the performance.
Note that optimization of $\lambda$ with
simulations is very time consuming, 
while it is easy to do even with brute-force 
search using Proposition~\ref{prop:perf_reconstruction}.

\section{Replica Analysis}
\label{sec:replicas}

In this section a sketch of derivation is given for
Propositions~\ref{prop:perf_reconstruction}~and~\ref{prop:mse}.
Throughout the rest of the paper, the replica trick given in 
Section~\ref{sec:statmech_form2} is assumed to be valid.
With this in mind, recall \eqref{eq:Z_replicated_1} and 
denote $\vm{v}^{a} = \vm{A}(\vm{x}^{0}- \vm{x}^{a})$.
The term inside $\ln$ in \eqref{eq:freeE_replica_real} 
can then be written as
\begin{IEEEeqnarray}{rCl}
\E_{\vm{x}^{0}} \bigg\{
\int
\prod_{a=1}^{\NR}
\Big[
\e^{- \beta \lambda \|\vm{x}^{a}\|_{1}}
\dx \vm{x}^{a}\Big]
\E_{\vm{A,\vm{w}}}
\prod_{a=1}^{\NR}
\e^{-\beta
\|\vm{v}^{a}  + \vm{w} \|_{1}}\bigg\},
\label{eq:replica_1}
\IEEEeqnarraynumspace
\end{IEEEeqnarray}
where
$\vm{x}^{0}$ has IID elements drawn according to
$p(x;\,\rho_{x},\sigma_{x}^{2})$.
We first concentrate on evaluating 
the latter term 
$\mathcal{I}_{\NR,\beta} (\setX) =
\E_{\vm{A},\vm{w}}
\prod_{a=1}^{\NR}
\e^{-\beta
\|\vm{v}^{a}  + \vm{w} \|_{1}}$,
for a fixed set 
$\setX= \{\vm{x}^{a}\}_{a=0}^{\NR}$.

Since $\vm{A}$ has IID elements 
with density $g_{A}(0,1/\N)$, conditioned on
$\setX$ the vectors $\{\vm{v}^{a}\}$
tend to jointly Gaussian RVs
by the central limit theorem 
as $\N \to \infty$.
More precisely,
if $\vm{v} \in \R^{\NR\N}$ is
formed by stacking $\{\vm{v}^{a}\}_{a=1}^{\NR}$
then $\vm{v}$ is a zero-mean Gaussian RV with 
covariance matrix $\vm{R} = \E_{\vm{A}} \vm{v}\vm{v}^{\trans}$.
We write this as $\vm{v}\sim g_{\vm{v}}(\vm{0},\vm{R})$
and remark that the $(a,b)$th ($a,b=1,\ldots,\NR$)
block of $\vm{R}$ is given by
\begin{IEEEeqnarray}{rCl}
\vm{R}_{a,b} 
  &=& 
  \big[Q_{00} 
  - \big(Q_{a0} + Q_{0b}\big) 
  + Q_{ab} \big]
  \vm{I}_{\M},
  \label{eq:Qab}
\end{IEEEeqnarray}
where
$Q_{ab} = \N^{-1}(\vm{x}^{a} \cdot \vm{x}^{b}).$ 
For later use, let the matrix $\vm{Q}\in\R^{(\NR+1)\times(\NR+1)}$ 
be composed of the elements $\{Q_{ab}\}$.
Thus, for large $\N$,
\begin{equation}
\mathcal{I}_{\NR,\beta} (\setX) =
\E_{\vm{w}}\int g_{\vm{v}}(\vm{0},\vm{R}) 
\exp \bigg[-\beta
\sum_{a=1}^{\NR}
\|\vm{v}^{a}  + \vm{w} \|_{1}\bigg]
\dx \vm{v},
\label{eq:I_1}
\end{equation}
where we have omitted terms that vanish as 
$\N\to\infty$ \cite{Tanaka-2002, Guo-Verdu-2005}.

In the large system limit of $N \to \infty$, Laplace's 
(the saddle point) method with respect to $\vm{R}$ yields the exact
assessment of $N^{-1} \ln E\{[Z_\beta (\vm{y},\vm{A};\lambda)]^u \}$ 
for $\forall{n}\in \mathbb{N}$ 
and $\forall{\beta} >0$. We here assume that the dominant saddle point
in the assessment is invariant under any permutation of the replica
indexes $a=1,2,\ldots,\NR$, which is often termed the replica symmetric
(RS) ansatz and is characterized as 
$Q_{a0} = Q_{0b} = m, 
Q_{aa} = Q, a=1,\ldots,\NR$ and $Q_{ab} = q$
for $a \neq b \in \{1,\ldots,\NR\}$ in the
current case. This allows us to express $\vm{v}^a$ in 
\eqref{eq:I_1} as
$\vm{v}^{a} =  \vm{z}_{a} \sqrt{Q-q}  + \vm{t} \sqrt{p-2 m + q} 
\in \R^{\M}$, which means that 
$\mathcal{I}_{\NR,\beta} (\setX)$
is proportional to
\begin{equation}
\Bigg[
\E\bigg\{
\bigg(\int 
\e^{-\beta
|z \sqrt{Q-q}  + t \sqrt{p-2 m + q} + w | 
- \frac{1}{2} z^{2}}
\dx z
\bigg)^{\NR}\bigg\}\Bigg]^{\M}\!\!\!\!, 
\label{eq:I_2}
\end{equation}
where 
$\E \{\,\cdots\} = \int (\,\cdots)
p(w;\,\rho_{w},\sigma_{w}^{2}) \dx w \Dx t$
and $\Dx t =  \dx t \e^{-t^{2}/2} /\sqrt{2\pi} $.
Since we are interested in the zero temperature solution
$\beta\to\infty$, Laplace's method
for the integral w.r.t.\ $z$ implies
\begin{IEEEeqnarray}{l}
\mathcal{I}_{\NR,\beta}
(\vm{Q} ) \propto
\Big[
\E
\e^{-\NR \beta
\psi(t, w; \vm{Q} )}\Big]^{\alpha\N}.
\label{eq:I_3}
\end{IEEEeqnarray}
We write next the exponential term in
\eqref{eq:I_2} in a slightly different form
by denoting $\chi = \beta(Q-q) \geq 0$
and $\varphi(t, w;\vm{Q}) =  
t \sqrt{p-2 m + Q} + w$.  
We also use the fact that 
$p-2 m + q = 
p-2 m + Q - \chi/\beta
\to p -2 m + Q$ for any finite $\chi$.
The Laplace's method requires then that
\begin{align}
\psi(t, w; \vm{Q} )
& = \min_{z} \bigg\{
|z \sqrt{\chi}  
+ \varphi(t, w;\vm{Q})
| 
+ \frac{z^{2}}{2}\bigg\}.
\end{align}
Examining the critical points of $\psi$ for a fixed set
$\{t, w, \chi, \vm{Q}\}$ 
shows that the minimizing $z$ gives
\begin{IEEEeqnarray}{l}
\psi(t, w; \vm{Q} ) = 
\begin{cases}
\varphi(t, w;\vm{Q})^{2} / (2\chi), & |\varphi(t, w;\vm{Q})|  < \chi; \\
|\varphi(t, w;\vm{Q})| - \chi / 2, & |\varphi(t, w;\vm{Q})|  > \chi.
\end{cases} \IEEEeqnarraynumspace
\end{IEEEeqnarray}

The next task is to average $\mathcal{I}_{\NR,\beta}
(w;\, \vm{Q} )$ over the set $\setX$.  The expectation
w.r.t.\ $\vm{Q}$ can be carried out under the 
RS ansatz by defining first the probability weight
\begin{IEEEeqnarray}{rCl}
\mu(\vm{Q}) &=&
\int\dx \vm{x}^{0} p(\vm{x}^{0})
\prod_{a=1}^{\NR}
\Big(\dx \vm{x}^{a}
\e^{-\beta \lambda \|\vm{x}^{a}\|_{1}}\Big) 
\IEEEnonumber\\
&&\; \times \prod_{0 \leq a\leq b \leq \NR}
\delta\big(\vm{x}^{a}\cdot\vm{x}^{b}-\N Q_{ab}\big),
\label{eq:Qweight} 
\end{IEEEeqnarray}
and integrating then w.r.t.\ the measure $\mu(\vm{Q})$.
Under the RS ansatz, measure \eqref{eq:Qweight} has the same form as 
in \cite{Kabashima-Wadayama-Tanaka-2009,
Kabashima-Vehkapera-Chatterjee-2012} so we 
skip the derivation here due to space constraints 
and arrive straight at the expression
\begin{IEEEeqnarray}{rCl}
\mu(\vm{Q})
&\propto& \int 
\dx \hat{\vm{Q}}
\exp\bigg[ \beta \N\bigg(
 \NR
 \frac{ 
 \hat{Q} Q
 - \chihat \chi}{2}
 - \NR m \mhat \IEEEnonumber\\
 && + \frac{\NR^{2}}{2}(\chihat \chi - \beta \chihat Q)
 + \frac{1}{\beta} \log \mathcal{M}_{\NR}
 (\hat{\vm{Q}};\,\beta, \lambda)
 \bigg)\bigg], \IEEEeqnarraynumspace
 \label{eq:muQ}
\end{IEEEeqnarray}
where $\hat{\vm{Q}}$ is a short-hand for 
$\{\chihat,\hat{Q},\mhat\}$.  We also have
the moment generating function for the elements of
$\{\vm{x}^{a}\}_{a=0}^{\NR}$
\begin{IEEEeqnarray}{rCl}
\mathcal{M}_{\NR}(\hat{\vm{Q}};\,\beta, \lambda) 
 &=&
 (1-\rho_{x})
  \E_{z}
   \e^{-\beta \NR \phi_{\lambda}(z\sqrt{\chihat};\, 
   \hat{Q})} 
   \IEEEnonumber\\
 && +
  \rho_{x}
  \E_{z}
   \e^{-\beta \NR 
   \phi_{\lambda}(z\sqrt{\chihat+\sigma_{x}^{2}\mhat^{2}};\, 
   \hat{Q})},  \IEEEeqnarraynumspace
\label{eq:MGF}
\end{IEEEeqnarray}
where $\E_{z} (\,\cdots)$ denotes 
$\int (\,\cdots) \Dx z $ and
$\phi$ satisfies
\begin{equation}
\label{eq:phi}
\phi_{\lambda} (h;\,\Qhat) = 
\begin{cases}
\vspace*{1ex}
-(|h|-\lambda)^{2}/(2 \Qhat), & \text{ if } |h| > \lambda \\
\qquad 0, & \text{ if } |h| \leq \lambda.
\end{cases}
\end{equation}
The final form of $\mu(\vm{Q})$ seems undoubtedly 
cryptic for a casual reader, so let us sketch the derivation
briefly (more details in 
\cite{Kabashima-Wadayama-Tanaka-2009,
Kabashima-Vehkapera-Chatterjee-2012}).  
The first task in obtaining 
\eqref{eq:muQ} is to write the Dirac's delta functions
using (inverse) Fourier transform and integrating over 
$\vm{x}^{0}$ with the help of the Gaussian integral
\begin{equation}
\sqrt{\frac{1}{2 \pi}}
\int \e^{-a x^{2} / 2 + b x} \dx x
= \frac{1}{\sqrt{a}} \exp\bigg(\frac{b^{2}}{2 a}\bigg).
\label{eq:gauss-int}
\end{equation}
Then \eqref{eq:gauss-int} is used right-to-left to 
decouple the replicated terms $\{\vm{x}^{a}\}$
and the average over them is obtained using the 
saddle point method 
as $\beta\to\infty$.  These last two steps give arise to 
\eqref{eq:phi} and the integrals in \eqref{eq:MGF}.
Rest of the terms in \eqref{eq:muQ} come essentially 
from the (inverse) 
Fourier transform of the Dirac's delta functions
where the hatted variables represent scaled
transform domain variables.

Combining \eqref{eq:I_3} and \eqref{eq:muQ} yields 
an expression for \eqref{eq:replica_1} as
\begin{IEEEeqnarray}{l}
\E\{[Z_{\beta}(\vm{y},\vm{A};  \lambda)]^{\NR}\} 
\propto  
\int \dx \vm{Q} 
\mathcal{I}_{\NR,\beta}(\vm{Q} )
\mu(\vm{Q}) \\
 = 
 \int   \dx \vm{Q} \dx \hat{\vm{Q}}
 \exp\Bigg\{ \beta \N\Bigg(
  \frac{\alpha}{\beta} \log 
  \E_{t,w} \e^{-\NR \beta\psi(t, w; \vm{Q} )}
 - \NR m \mhat \IEEEnonumber\\
  + \NR
  \frac{ 
  \hat{Q} Q
  - \chihat \chi}{2}
 + \frac{\NR^{2}}{2}(\chihat \chi - \beta \chihat Q)
   + \! \frac{1}{\beta} \log \mathcal{M}_{\NR}(\hat{\vm{Q}};\,\beta, \lambda)  \Bigg) \!\Bigg\}. \IEEEnonumber
\end{IEEEeqnarray} 
For the integration w.r.t.\ $\vm{Q}$ and $\hat{\vm{Q}}$ 
we use again the saddle point method as $\N\to\infty$.  Note 
that we have then by the law of large numbers 
$p \to \sigma_{x}^{2} \rho_{x}$ as well (see \eqref{eq:I_2}).
Thus, the \emph{replica symmetric} 
expression for \eqref{eq:freeE_replica_real} reads
\begin{IEEEeqnarray}{l}
f_{\mathsf{rs}}(\lambda) =
 - \extr\Bigg\{
  \frac{ \hat{Q} Q}{2} - 
  \frac{\chihat \chi}{2}
 - m \mhat \\
  +\!\lim_{\NR\to 0^{+}}\!
   \frac{\partial}{\partial \NR} \bigg[
    \frac{\alpha}{\beta} 
    \log \E \e^{-\NR \beta\psi(t, w; \vm{Q} )}
   \!+\! \frac{1}{\beta} 
   \log \mathcal{M}_{\NR}(\hat{\vm{Q}};\,\beta, \lambda)
   \bigg]\!
   \Bigg\}, \IEEEnonumber
\end{IEEEeqnarray}
where we used the fact that the 
order of extremization $\extr\{\,\cdots\}$ w.r.t.\
$\{\chi, m, Q,\chihat,\mhat,\hat{Q}\}$
and the partial derivative
w.r.t.\ $u$ can be exchanged \cite{Tanaka-2002}.
Solving the remaining derivatives finally gives 
the form
\begin{IEEEeqnarray}{l}
f_{\mathsf{rs}}(\lambda) =
\extr
\bigg\{
  \frac{\chihat \chi-\hat{Q} Q}{2}
 + m \mhat 
 + \alpha \E_{t,w}  \psi(t, w; \chi, m, Q )  \nonumber\\
  +  (1-\rho_{x}) \E_{z}
      \phi_{\lambda}\big(z\sqrt{\chihat};
     \hat{Q}\big) 
  + \rho_{x} \E_{z}
     \phi_{\lambda}
     \big(z\sqrt{\chihat+\sigma_{x}^{2}\mhat^{2}};
     \hat{Q} \big) \!
\Big\} \IEEEnonumber\\
 \label{eq:freeE_final2}
\end{IEEEeqnarray}
in the limit $\NR \to 0^{+}$. 

We have now managed to write the normalized 
free energy under RS ansatz as the solution of an 
extremization problem that has a couple of 
expectations inside.  Let us first consider 
the derivatives 
w.r.t.\ the variables
$\{\chihat,\mhat,\hat{Q}\}$.
Since $\psi$ does not depend on them, 
we only need to solve the expectations and partial derivatives 
on the second line in \eqref{eq:freeE_final2}.

\begin{lemma}
Let $h$ be a real positive (function) independent of $z\in \R$.
Then, for positive real parameters $\Qhat$ and $\lambda$
we have 
\begin{IEEEeqnarray}{rCl}
\E_{z} \phi_{\lambda}
  (z \sqrt{h};\,\Qhat) &=&
  \Qhat^{-1} r_{ \lambda}(h), \\
\frac{\partial}{\partial x} r_{ \lambda}(h)
&=& - \bigg(\frac{\partial h}{\partial x}\bigg)
Q\bigg(\frac{\lambda}{\sqrt{h}}\bigg),
\end{IEEEeqnarray}
where $r_{\lambda}(h)$ is given in \eqref{eq:rfunc_prop}.
\end{lemma}

Using the above results, the normalized free energy reads
\begin{IEEEeqnarray}{l}
f_{\mathsf{rs}}(\lambda) =
\extr
\bigg\{
 \frac{ \hat{Q} Q}{2} - 
  \frac{\chihat \chi}{2}
 - m \mhat - \alpha \E_{t,w}  \psi(t, w; \chi, m, Q )  \nonumber\\
 \qquad\quad + \Qhat^{-1} \big[
  (1-\rho_{x}) r_{\lambda}(\chihat)
+ \rho_{x} r_{\lambda}(
\chihat+\sigma_{x}^{2}\mhat^{2}) \big]
\big\},
 \label{eq:freeE_final3}
\end{IEEEeqnarray}
where $\chi$ is given in \eqref{eq:chi_prop} and
\begin{IEEEeqnarray}{rCl}
\label{eq:m}
m &=& 
2\sigma^{2}_{x}\rho_{x} \bigg(\frac{\mhat}{{\Qhat}}\bigg)
Q\bigg(\frac{\lambda}{\sqrt{\chihat+\sigma_{x}^{2}\mhat^{2}}}\bigg), \\
\label{eq:Q}
Q &=&
 - 2\Qhat^{-2} \big[
  (1-\rho_{x}) r_{\lambda}(\chihat)
+ \rho_{x} 
r_{\lambda}(\chihat+\sigma_{x}^{2}\mhat^{2}) \big].
\IEEEeqnarraynumspace
\end{IEEEeqnarray}
To obtain rest of the parameters, we need the following result.

\begin{lemma}
Let $\omega(t\sqrt{a},x_{1},\ldots,x_{k})$ be a real-valued 
function, where $a \geq 0$ and 
$\{t,x_{1},\ldots,x_{k}\}$ are independent 
random variables that do not depend on $a$. Then,
\begin{equation}
\frac{\partial}{\partial a}
\int\! \omega(t\sqrt{a},x_{1},\ldots,x_{k}) \Dx t
\!=\! \frac{1}{2}\int\! \omega''(t\sqrt{a},x_{1},\ldots,x_{k}) \Dx t,
\label{eq:f_diff_1}
\end{equation}
where $\omega''(\,\cdots)$ is the 2nd order partial derivative 
w.r.t.\ first argument.
Also, denoting the indicator function $\indF\{\,\cdots\}$,
\begin{IEEEeqnarray}{rCl}
\label{eq:Id1}
\int \indF\{|t| > a\}\Dx t &=& 2 Q(a), \\
\label{eq:Id3}
\int t^{2}\indF\{|t| < a\}\Dx t &=& 
1 - 2 Q(a)
    - \frac{2 a}{\sqrt{2\pi}} \e^{-a^{2}/2},
    \IEEEeqnarraynumspace
\end{IEEEeqnarray}
where the integrals are over the set of real numbers.
\end{lemma}

Using \eqref{eq:f_diff_1} for the partial 
derivatives w.r.t.\  $m$ and $Q$, and then 
\eqref{eq:Id1}~--~\eqref{eq:Id3}
for the remaining integrals shows that
$\hat{Q} = \hat{m}$ as given in \eqref{eq:hatQ-hatm_prop}.
Furthermore, $\mse = \sigma_{x}^{2} \rho_{x} - 2 m + Q$
reduces to \eqref{eq:mse_prop_1} and 
gives the MSE of the reconstruction 
\cite{Kabashima-Wadayama-Tanaka-2009,
Kabashima-Vehkapera-Chatterjee-2012}.
Similarly,
from the derivative of $\chi$ 
and \eqref{eq:Id1}~--~\eqref{eq:Id3} one gets
\eqref{eq:chihat_prop}.
Thus, we have obtained a full description of the 
free energy under the RS ansatz in terms of six 
parameters.  More importantly, 
we obtained as a by product the MSE behavior 
of the convex optimization problem based on 
\eqref{eq:CS_Standard_Problem_LASSO} and
\eqref{eq:cost_l1}, finishing the proof 
of Proposition~\ref{prop:mse}.

To obtain Proposition~\ref{prop:perf_reconstruction},
we require that $\mse \to 0$.  This implies
$\rho_{x}\sigma_{x}^{2} = m = Q$ and
$\hat{m} = \hat{Q} \to \infty \implies \chi \to 0$.
For a non-trivial solution we also need 
$\hat{\chi} \in O(1)$ and $0<\lambda<\infty$.
However, the condition for critical 
threshold $\alpha$ cannot be directly obtained
by plugging this to
\eqref{eq:mse_prop_1}~--~\eqref{eq:chihat_prop}.
Instead, we expand the Q-function
and exponential function near zero
with the Taylor series, define $\kappa = \mse / \chi^{2}$
and examine the limits for $\kappa$ and $\hat{\chi}$.
Some algebra provides then Proposition~\ref{prop:perf_reconstruction}.

\bibliography{./jour_short,./conf_short,./biblio_saikat_CS}

\bibliographystyle{IEEEtran}

\end{document}